\documentclass[twocolumn,secnumarabic,amssymb, nobibnotes, aps, pra]{revtex4-1}
\usepackage{amsfonts}
\usepackage{graphicx}
\usepackage{epsfig}
\usepackage{epsf}
\usepackage{amssymb}
\usepackage{amsmath}
\usepackage{amsthm}
\usepackage{multirow}
\usepackage[usenames, dvipsnames,table,xcdraw]{xcolor}
\usepackage{mathrsfs}
\usepackage{bm}
\usepackage{dcolumn}
\usepackage{epstopdf}
\usepackage{bm,bbm}
\usepackage{cases,booktabs}
\usepackage[a4paper, pagebackref=false, bookmarks=false, colorlinks=true,
linkcolor=blue, citecolor=blue,
pdfauthor={ },
pdftitle={ },
pdfsubject={ },
pdfkeywords={ }]{hyperref}

\newcommand{\ket}[1]{\vert#1\rangle}
\newcommand{\bra}[1]{\langle#1\vert}

\makeatletter
\newcommand\figcaption{\def\@captype{figure}\caption}
\newcommand\tabcaption{\def\@captype{table}\caption}
\makeatother

\begin{document}

\title{Cooling Flexural Modes of a Mechanical Oscillator by Magnetic Trapped Bose-Einstein Condensate Atoms}

\author{Donghong Xu$^{1,2,3}$}
\author{Fei Xue$^{1,3}$}
\email[]{xuef@hmfl.ac.cn}

\affiliation{
$^{1}$Anhui Province Key Laboratory of Condensed Matter Physics at Extreme Conditions, High Magnetic Field Laboratory, Chinese Academy of Sciences, Hefei 230031, Anhui, People's Republic of China\\
$^{2}$University of Science and Technology of China, Hefei 230026, People's Republic of China\\
$^{3}$Collaborative Innovation Center of Advanced Microstructures, Nanjing University, Nanjing 210093, People's Republic of China}

\date{\today}

\begin{abstract}
We theoretically study cooling of flexural modes of a mechanical oscillator by Bose-Einstein Condensate (BEC) atoms (Rb87) trapped in a magnetic trap. The mechanical oscillator with a tiny magnet attached on one of its free ends produces an oscillating magnetic field. When its oscillating frequency matches certain hyperfine Zeeman energy of Rb87 atoms, the trapped BEC atoms are coupled out of the magnetic trap by the mechanical oscillator, flying away from the trap with stolen energy from the mechanical oscillator. Thus the mode temperature of the mechanical oscillator is reduced. The mode temperature of the steady state of the mechanical oscillator, measured by the mean steady-state phonon number in the flexural mode of the mechanical oscillator, is analyzed. It is found that ground state (phonon number less than 1) may be accessible with optimal parameters of the hybrid system of a mechanical oscillator and trapped BEC atoms.
\end{abstract}

\maketitle

\section{Introduction}

Mechanical oscillators play an important role in applications of ultra-high precise measurements and sensing of displacements \cite{Moser2013}, masses \cite{Chaste2012} and nuclear spin polarization fluctuations \cite{Xue2011,Peddibhotla2013}. Mechanical oscillators are also expected to play an important role in exploring various meso-scopic quantum phenomena. They have been used in study of fundamental quantum physics, such as generating entangled states and deterministic entanglement \cite{Cleland2004, Wang2011}. Mechanical oscillators could serve as a quantum data bus in quantum computations \cite{Xue2007a,Rabl2010}. Reducing thermal vibrations of mechanical oscillators is important in their practical applications and fundamental physics studies of micro and nano-mechanical oscillators. Therefore, great efforts are done in cooling down the thermal vibrations of mechanical oscillator to a temperature of quantum regime both in theoretical and experimental research. On the experimental side, although passive cooling of a mechanical oscillator down to quantum ground state has been demonstrated in recent experiment by using a microwave-frequency mechanical oscillator of gigahertz frequency (about 6 GHz), using standard dilution refrigerator (20mK) \cite{Connell2010}. For mechanical oscillators with oscillating frequencies at MHz, the passive cooling by dilution refrigerator is hard.   Significant effort has been devoted to developing active cooling methods in hybrid system, such as cooling mechanical oscillator by optical field enhanced by a cavity \cite{Aspelmeyer2014}, by single electron transistor \cite{Naik2006}, or by coupling it to a superconduction LC circuit \cite{Teufel2008}. Proposals suggest that the ground state cooling of mechanical oscillator is possible \cite{Tian2009, Xue2007b, Marquardt2007}. Recently, the hybrid system cooling evokes some attentions as an intriguing cooling proposal for mechanical oscillators, such as atom-assisted cooling in opto-mechanics system. Specially, the atom based hybrid system was proposed few years ago \cite{Treutlein2007,Deng2014}. Recently, by applying the Landau-Zener theory, a cooling scheme in such hybrid system, is proposed \cite{Tretiakov2016}.

Coupling between mechanical oscillators and alkali atoms $^{87}Rb$ in gas state via Hyperfine Zeeman splitting of these atoms at room temperature was experimentally demonstrated \cite{Wang2006}. A few years ago, interactions between a mechanical oscillator and Bose-Einstein-Condensation (BEC) atoms via surface force of the mechanical oscillator was realized \cite{Hunger2010}. Recently, magnetic cantilever tip resonant with trapped cold atoms is presented \cite{Montoya2015}. The magnetic trapped BEC \cite{Leggett2001} and mechanical oscillators on atom chips are reviewed in Ref. \cite{Hunger2011}. In this work, we study the cooling of flexural modes of a mechanical oscillator by coupling them to Bose-Einstein Condensate atoms ($^{87}Rb$). These atoms are trapped in a magnetic trap. A magnetic tip is attached on the free end of mechanical oscillator. The tiny magnetic tip produces an oscillating magnetic field around the center of the magnetic trap while the mechanical oscillator vibrates. This oscillating magnetic field induces trapped atoms' spin flip. These `flipped' atoms then escape from the magnetic trap with stolen energy from the mechanical oscillator. These atoms are excited to untrappable states, then freely expand, get away from the trap \cite{Mewes1997, Steck1998}.

The paper is organized as follows. In Sec. \ref{hybrid system}, we introduce the hybrid system and its Hamiltonian. For mechanical oscillator coupling to a single atom in the trap, the Hamiltonian of this hybrid system can be described with the Jaynes-Cummings model. For mechanical oscillator coupling to an ensemble of cold atoms undistinguishably, the coupling constant between mechanical oscillator and the atoms is enhanced \cite{Treutlein2007,Tretiakov2016,Patton2013}. The hybrid system is described by Tavis-Cummings Hamiltonian. In Sec. \ref{Expression of steady phonon number after cooling}, we draw an analogy between the excited atoms and the output of atom laser, then the master equation for phonon number in mechanical oscillator's fundamental mode is derived. The expression of mean steady-state phonon number is obtained. In Sec. \ref{discussion and conclusion}, by applying some practical parameters, it is found that the mechanical oscillator could be cooled down to its ground state.

\section{Hybrid system of a mechanical oscillator and BEC atoms}
\label{hybrid system}

\begin{figure}[tp]
%\centering
\includegraphics[width=8cm]{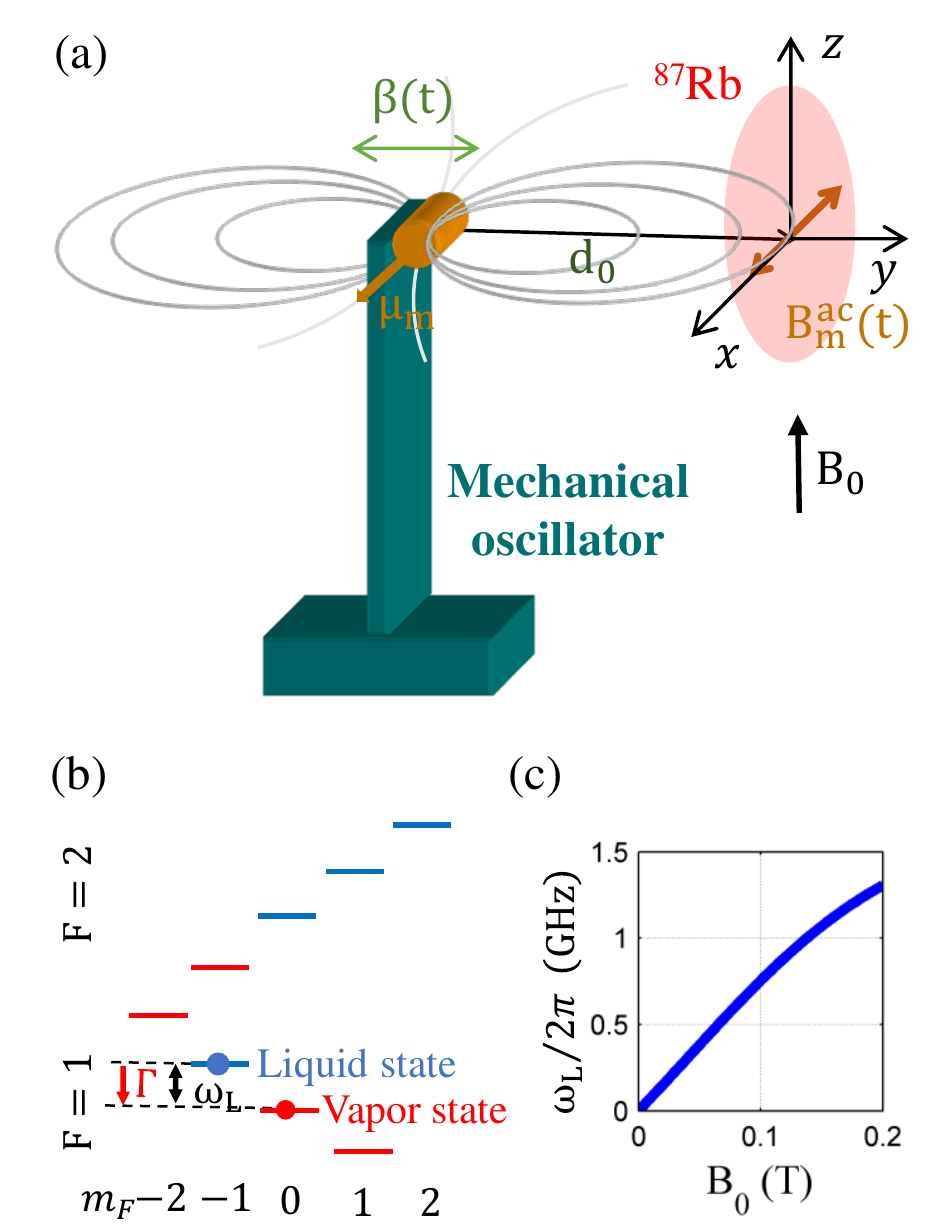}
\begin{flushleft}
\caption{(Color online). (a) Schematic layout for the hybrid system. The magnetic trapped BEC atoms is in red color and the mechanical oscillator is in green color. The tiny Cobalt magnetic tip (in yellow color) attached on the free end of the mechanical oscillator produces an oscillating magnetic field around center of the trap while the mechanical oscillator vibrates. The single-domain magnetic tip creates a field $\boldsymbol{\textbf{B}}_{\text{m}}^{\text{ac}}(t)$ and couples to atomic spin of atoms. (b) Hyperfine structure level of $^{87}Rb$ in the static magnetic field $\textbf{B}_{0}$. The trappable states are in blue color and the untrappable states are in red color. The hyperfine level $\ket{1,-1}$ are coupled out to the mechanical oscillator when the Larmor frequency $\omega_{\text{L}}$ is tuned to the oscillation frequency of mechanical oscillator $\omega_{\text{m}}$, and then, atoms are excited from $\ket{1,-1}$ to $\ket{1,0}$ with a transition rate $\Gamma$. (c) Field dependence of Larmor frequency $\omega_{\text{L}}/2\pi\,=\,(E_{\ket{1,-1}}-E_{\ket{1,0}})/h$ between state $\ket{1,-1}$ (liquid state) and $\ket{1,0}$ (vapor state).}
\label{fig:hybridsystem}
\end{flushleft}
\end{figure}

The hybrid system is illustrated in Fig. \ref{fig:hybridsystem}(a). In this work, the atoms are trapped in a magnetic trap. The center of trap is at a distance $d_{0}$ from the mechanical oscillator along the y-axis. A single-domain magnetic tip creates a magnetic field with gradient $G_{\text{m}}$ while the mechanical oscillator vibrates. The mechanical oscillator oscillates along the y-axis: $\beta(t)\,=\,\beta \cos(\omega_{\text{m}}t\,+\,\varphi_{\text{m}})$. Here, $\beta$ is the amplitude of vibrations, $\varphi_{\text{m}}$ is the phase and $\omega_{\text{m}}$ is the frequency of mechanical oscillator. We choose the orientation of the Cobalt magnet tip so that $\boldsymbol{\textbf{B}}_{\text{m}}^{ac}(t)$ is perpendicular to the static magnetic field $\boldsymbol{\textbf{B}_{0}}$ around center of the trap. Because of the shape anisotropy of tip, magnetic moment $\boldsymbol{\mu}_{\text{m}}$ is spontaneously oriented along its long axis, as shown in Fig. \ref{fig:hybridsystem}(a), in yellow color.  Let the static magnetic properly set along the z-axis:
$\boldsymbol{\textbf{B}_{0}}\,=\,B\boldsymbol{e_{\text{z}}}$
with $B$ is the amplitude of the applied static magnetic field. Approximating the magnetic tip as magnetic dipole, we have the field gradient which is similar to \cite{Treutlein2007,Tretiakov2016,Steinke2011}:
$G_{\text{m}}\,=\,3\mu_{0}\left|\boldsymbol{\mu}_{\text{m}}\right|/(4\pi d_{0}^{4})$
around center of the trap. $\mu_{0}$ is the permeability of vacuum. Assuming the oscillating amplitude $\beta\, \ll \, d_{0}$ thus the mechanic motion of the mechanical oscillator induces an oscillating magnetic field in the x-direction as illustrated in Fig. \ref{fig:hybridsystem}(a):
\begin{equation}
\boldsymbol{\textbf{B}}_{\text{m}}^{\text{ac}}(t)=G_{\text{m}}\beta(t)\boldsymbol{e_{\text{x}}}.
\end{equation}

For BEC [Fig. \ref{fig:hybridsystem}(a), in red color], the static magnetic field $\boldsymbol{\textbf{B}_{0}}$ is applied along the z-axis on a group of $^{87}Rb$ atoms, separating their degenerate hyperfine levels: $\boldsymbol{F}\,=\,\boldsymbol{I}\,+\,\boldsymbol{J}$. Here, the electronic angular momentum $J\,=\,1/2$, the nuclear spin $I\,=\,3/2$. Denoting the hyperfine levels as $\ket{F,m_{\text{F}}}$, with $m_{\text{F}}$ being the projection of $\boldsymbol{F}$ on the axis of the applied static field $\boldsymbol{\textbf{B}_{0}}$, the hyperfine Zeeman levels depend on the amplitude $B\,\equiv\,\left|\boldsymbol{\textbf{B}_{0}}\right|$ of the applied magnetic field in a special case of Breit-Rabi formula \cite{Leggett2001}:
\begin{equation}
E_{\ket{F,m_{\text{F}}}}=(-1)^{\text{F}}\frac{A_{\text{hf}}\,h}{2}\sqrt{1+m_{\text{F}}\frac{B}{B_{\text{hf}}}+(\frac{B}{B_{\text{hf}}})^2}.
\label{Breit-Rabi}
\end{equation}
In Eq. (\ref{Breit-Rabi}), we have offset the zero of energy to the mean of zero field energies of the atomic spin states F\,=\,1 and F\,=\,2 for convenience. The zero-field energy splitting $A_{\text{hf}}\,=\,(E_{\text{F}\,=\,2}-E_{\text{F}\,=\,1})/h\,\approx\,6.835$ GHz, and $B_{\text{hf}}\,\equiv\,A_{\text{hf}}\,h/(2\mu_{B})\approx\,$ 0.24 T is called characteristic hyperfine crossover field \cite{Leggett2001}. Here, $h$ is the Planck constant, $\mu_{B}$ is the Bohr magneton.

We consider a trap potential: $V_{T}(\text{x},\text{y},\text{z})\,=\,(m/2)(\omega_{\text{x}}^2\text{x}^{2}+\omega_{\text{y}}^{2}\text{y}^{2}+\omega_{\text{z}}^{2}\text{z}^{2})$ \cite{Darazs2014}. Here, m is the atomic mass of $^{87}Rb$, $\omega_{\text{x}}\,=\,\omega_{\text{y}}$ is transversal trapping frequency, $\omega_{\text{z}}$ is the longitudinal trapping frequency. For such a pure magnetic trap, the states $\ket{2,2}, \ket{2,1}, \ket{2,0}$ and $\ket{1,-1}$ are attracted to the local minimum of the magnetic trap, while the states $\ket{2,-1}, \ket{2,-2}, \ket{1,1}$ and $\ket{1,0}$ are free out of the magnetic trap. These eight hyperfine spin levels of trappable (in blue color) and untrappable (in red color) and the process of atoms' transition for $^{87}Rb$ induced by the mechanical oscillator are shown in Fig. \ref{fig:hybridsystem}(b).

Because the other two trappable states $\ket{2,1}$ and $\ket{2,0}$ tend to change into untrappable states by two-body collision among atoms \cite{Leggett2001} and the collision loss is lower in F\,=\,1 than F\,=\,2 \cite{Treutlein2007}, the cold atoms in the trap are prepared in state $\ket{1,-1}$ initially. We choose $\ket{1,-1}\,\equiv\,\ket{0}$, named as liquid state. We choose $\ket{1,0}\,\equiv\,\ket{1}$, named as vapor state.

\textit{Jaynes-Cummings Model.} We assume that all BEC state atoms are prepared in the liquid state $\ket{0}$. Each atom changing from the liquid state $\ket{0}$ to the vapor state $\ket{1}$ is expected to taken energy from the mechanical oscillator. The energy transfer rate depends on atoms' ``transition rate''. The trapped BEC atoms in the liquid state are assumed to be in the Thomas-Fermi (TF)regime. For $B\,\ll\, B_{\text{hf}}$, which is reliable in most BEC experiments \cite{Leggett2001}, the Larmor frequency between the liquid state and the vapor state can be written as:
\begin{equation}
\omega_{\text{L}}=\frac{\mu_{B} \left|g_{\text{F}}\right|B}{\hbar},
\end{equation}
to the first order of $B$, which is plotted in Fig. \ref{fig:hybridsystem}(c).
Here, $g_{\text{F}}$ is the Land$\acute{e}$ g-factor of atomic spin state $\boldsymbol{F}$. As $g_{I}\,\ll \,g_{J}$, $g_{\text{F}}$ can be expressed by: $g_{\text{F}}\,=\,g_{J}[F(F\,+\,1)\,+\,J(J\,+\,1)\,-\,I(I\,+\,1)]/[2F(F\,+\,1)] $\cite{Woodgate1980}. $g_{J}$ and $g_{I}$ are the Land$\acute{e}$ g-factors for total angular momentum of electron and nuclear spin, respectively. Therefore, the Larmor frequency $\omega_{\text{L}}$ can be tuned by adjusting the amplitude of static field $B$, hence the detuning $\delta\,=\,\omega_{\text{m}}\,-\,\omega_{\text{L}}$ can also be tuned.

The oscillating magnetic field $\boldsymbol{\textbf{B}}_{\text{m}}^{\text{ac}}(\text{t})$ interacts with the liquid state and vapour state via the Zeeman Hamiltonian:
$H_{\text{z}}\,=\,\boldsymbol{-\mu\cdot \textbf{B}}^{\text{ac}}_{\text{m}}(t)$
with $\boldsymbol{\mu}\,=\,-\mu_{B}g_{\text{F}}\boldsymbol{F}$ \cite{Treutlein2007}. In the case of that the direction of field gradient at the center of the trap is along x-axis, as shown in Fig. \ref{fig:hybridsystem}(a), the Hamiltonian $H_{\text{z}}$ reads:
\begin{equation}
H_{\text{z}}=\mu_{B}g_{\text{F}}F_{\text{x}}G_{\text{m}}\beta(t).
\label{interaction}
\end{equation}

We replace the mechanical oscillator's displacement $\beta(t)$  by the operators: $\beta(t)$ $\rightarrow$ $a_{\text{qm}}(a\,+\,a^{\dagger})$ with $a_{\text{qm}}\,\equiv\,\sqrt{\hbar /2m_{\text{eff}}\omega_{\text{m}}}$ being the rms amplitude of the mechanical oscillator's quantum zero-point fluctuation, $m_{\text{eff}}$ being the effective mass of the oscillator, and $a$ ($a^{\dagger}$) being the bosonic annihilation (creation) operator for phonons in the flexural mechanical mode.

The two states of liquid state $\ket{0}$ and vapor state $\ket{1}$ can be separated from the other six hyperfine sub-levels by the quadratic Zeeman effect \cite{Treutlein2007,Tretiakov2016,Treutlein2006}. Therefore, we could define a pseudo spin with the liquid state and vapour state as its eigenstate of $\sigma_z$.
with $\sigma_\text{z}$ is the z component of Pauli matrix. With $\sqrt{2}g_{F}F_{\text{x}}$ replacing by $\sigma_{\text{x}}$ and applying the rotating-wave approximation in Eq. (\ref{interaction}), the interaction Hamiltonian $H_{I}$ of the mechanical oscillator interacting with the two-level system can be written as \cite{Treutlein2007,Tretiakov2016,Geraci2009}:
\begin{equation}
H_{I}=\frac{1}{2}\hbar g_{0}(a^{\dagger}\sigma_{-}+a\sigma_{+}).
\end{equation}
Here, $g_{0}\,=\,\mu_{B}G_{\text{m}}a_{\text{qm}}/(\sqrt{8}\hbar)$ is the single-atom-single-phonon coupling constant and $\sigma_{\pm}\,=\,\frac{1}{2}(\sigma_{\text{x}}\,\pm \,i\sigma_{\text{y}})$. Thus, we have the Hamiltonian
for the hybrid system of the mechanical oscillator and a single $^{87}Rb$ atom in the Jaynes-Cummings form:
\begin{align}
H_{JC}=\hbar\omega_{\text{m}}a^{\dagger}a+\frac{1}{2}\hbar\omega_{\text{L}}\sigma_{\text{z}}+\frac{1}{2}\hbar g_{0}(a^{\dagger}\sigma_{-}+a\sigma_{+}).
\label{JC-type}
\end{align}

\textit{Tavis-Cummings Model.} For $^{87}Rb$ BEC atoms, the average distance between them is much smaller than their typical wave length in the trap potential \cite{Leggett2006}, many $^{87}Rb$ are interacting with the mechanical oscillator simultaneously. Assuming that N $^{87}Rb$ atoms identically interact with the mechanical oscillator, the hybrid system is better described by the Tavis-Cummings Model. Thus, we define the following collective spin operators as \cite{Tretiakov2016,Shore1993,Xichen2015}:
$S_{\text{z}}=\sum_{i=1}^{N}\sigma_{\text{z}}^{i}$, $S_{\pm}=\frac{1}{\sqrt{N}}\sum_{i=1}^{N}\sigma_{\pm}^{i}$, where $i$ is the $ith$ atom and the sums run over the N atoms. These BEC state atoms are all identical particles that we do not distinguish them from each other. In view of that, we neglect the space distribution of these magnetic trapped cold atoms and assume the energy level spacing between the liquid state $\ket{0}$ and the vapor state $\ket{1}$ are all equal to $\hbar\omega_{\text{L}}$ \cite{Treutlein2007}, so that the Hamiltonian of these N two-level systems reads:
$\hbar\omega_{\text{L}}S_{\text{z}}/2$.
The interaction between mechanical oscillator and this ensemble of pseudo spins can be described as:
$g_{0}\sqrt{N}/2(a^{\dagger}S_{-}\,+\,aS_{+}).$
The coupling strength is enhanced \cite{ Patton2013}:
$g_{0}\sqrt{N}\rightarrow g_{N}$,
with $g_{N}$ being the collective coupling constant. The hybrid system, with N two-level atoms identically coupling to mechanical oscillator, is thus described by the Tavis-Cummings Hamiltonian \cite{Treutlein2007,Shore1993}:
\begin{equation}
H_{TC}=\hbar\omega_{\text{m}}a^{\dagger}a+\frac{1}{2}\hbar\omega_{\text{L}}S_{\text{z}}+\frac{1}{2}\hbar g_{N}(a^{\dagger}S_{-}+aS_{+}).
\label{TC-type}
\end{equation}
The Tavis-Cummings Hamiltonian describes the identical coupling between the BEC state atoms and mechanical oscillator, the coupling constant is collectively enhanced. In the following, we calculate the mean steady-state phonon number for the mechanical oscillator when it interacts with an ensemble of $^{87}Rb$ atoms.

\section{Mean steady-state phonon for mechanical oscillator}
\label{Expression of steady phonon number after cooling}

\begin{figure}[tp]
\centering
\includegraphics[width=7cm]{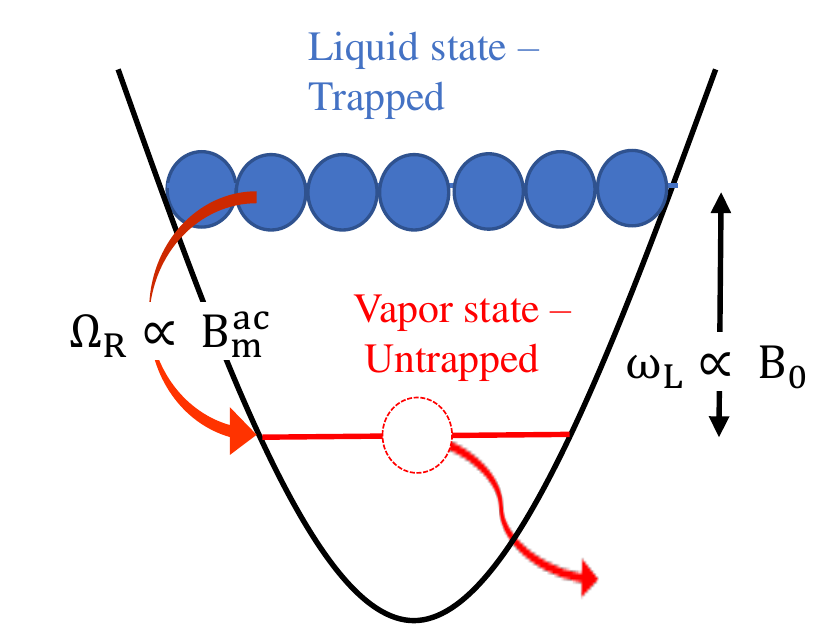}
\caption{(Color online). The atoms in the liquid state are transferred into the vapor state under the action of the magnetic field ${\textbf{B}}^{\text{ac}}_{\text{m}}(t)$ produced by the vibrating mechanical oscillator in half a Rabi cycle. When the atoms are excited to the vapor state, they get away from the trap with the energy from mechanical oscillator. }
\label{fig.2}
\end{figure}

The magnetic field produced by mechanical oscillator induces the liquid state been coupled to the vapor state with the Rabi frequency $\Omega_{\text{R}}$ \cite{Tretiakov2016}:
\begin{eqnarray}
\Omega_{\text{R}}=\frac{\mu_{B}G_{\text{m}} a_{\text{qm}} <(a+a^{\dagger})>}{\sqrt{8}\hbar},
\label{rabi}
\end{eqnarray}
with $<a_{\text{qm}}(a+a^{\dagger})>$ being thermal fluctuation of mechanical oscillator. The untrapped motional states in the atoms has a continuum  energy width which is given by $\mu_{c}$. This continuum energy width is BEC's chemical potential \cite{Treutlein2007,Leggett2006}. We here note the rate of BEC state atoms changed from liquid state to the vapor state as $\Gamma$. In the following, we will call $\Gamma$ as atoms' transition rate and the transition rate $\Gamma$ is expected to be derived according to Ref. \cite{Treutlein2007,Steck1998}. Neglecting the gravity, the transition rate of atoms from the liquid state $\ket{0}$ to the vapor state $\ket{1}$ reads:
\begin{equation}
\Gamma(\delta)=\zeta(\delta)\Omega^{2}_{\text{R}}.
\label{rate}
\end{equation}
Here, $\zeta(\delta)\equiv 15\pi \hbar/8\mu_{c}[\sqrt{\hbar\delta/\mu_{c}}-(\sqrt{\hbar\delta/\mu_{c}})^{3}]$, is related with the chemical potential $\mu_{c}$ and the detuning $\delta$.

The transition rate is negligible outside the resonance shell and strongly enhanced within the shell \cite{Steck1998}, and the mechanical oscillator resonance with the main axes $r_{i}\,=\,R_{i}\sqrt{\hbar\delta/\mu_{c}}$, where $R_{i}$ is the Thomas-Fermi radii of the BEC. When the atoms in resonance region are excited into the vapor state, other condensate outside the resonance shell will move into the resonance area, replacing the leaving ones \cite{Steck1998}.

In our case, the mechanical oscillator loses its energy, and induces the atoms' spin flips while it interacts with the BEC atoms. The atoms become untrapped and fly away. The atoms in the resonance shell are excited, changed from the liquid state to the vapor state. Note the interacting time between the mechanical oscillator and the atoms be $\tau$. When atoms are excited, they freely expand and fly away from the trap with the stolen energy from mechanical oscillator. Those atoms flying away from the trap play an important role in our cooling method. The interaction time can be approximated by half a Rabi cycle time:
\begin{equation}
\tau=\frac{\pi}{\Omega_{\text{R}}}.
\label{time}
\end{equation}

Following the master equation approach for the quantum laser theory in Ref. \cite{Orszag2016}, we derived the means steady-state phonon number of the mechanical oscillator. We assume that at time $t_{0}=0$, the mechanical oscillator and magnetic trapped BEC atoms do not interact each other, and define $\rho(t)$ as the reduced density operator of the mechanical oscillator at time $t$. When the interaction is switched on at time $t_{0}$, the atom is excited to the vapor state  from a liquid state atom. $\rho(t_{0}+\tau)$ is obtained by the super-operator $\boldsymbol{M}$ \cite{Orszag2016,Scully1997}:
\begin{equation}
\rho(t_{0}+\tau)=\boldsymbol{M}(\tau)\rho(t_{0}),
\end{equation}
in which:
\begin{equation}
\begin{split}
\boldsymbol{M}(\tau)\rho(t_{0}) & =\text{Tr}_{\text{BEC}}[e^{\frac{-iH_{I}\tau}{\hbar}}\rho(t_{0})\otimes\ket{0}\bra{0}e^{\frac{iH_{I}\tau}{\hbar}}].
\label{trace}
\end{split}
\end{equation}
Here, $H_{I}\,=\,\frac{1}{2}\hbar g_{0}(a\sigma_{+}\,+\,H.c)$ being the Jaynes-Cummings Hamiltonian in Eq. (\ref{JC-type}) at resonance in the interaction picture, and $\tau$ is the interaction time between the oscillator and the atoms. $\text{Tr}_{\text{BEC}}$ denotes tracing over the variables of the two-level system of this BEC state atoms.

We assume that the BEC atoms are identically coupled to the mechanical oscillator. When $k$ atoms are excited:
\begin{equation}
\rho(t)=\boldsymbol{M}^{k}(\tau)\rho(0).
\label{Eqevolution}
\end{equation}
Here, $k=\Gamma t$ is the number of excited atoms. With the presence of the dissipation mechanical oscillator noted as $\mathscr{L}[\rho]$, differentiating Eq. (\ref{Eqevolution}), we get the evolution of the reduced density operators $\rho$:
\begin{equation}
\begin{split}
\frac{d \rho}{dt}&= \Gamma[\ln\boldsymbol{M}(\tau)]\rho+ \mathscr{L}[\rho].
\label{Eqfirst}
\end{split}
\end {equation}

Since a single BEC atom takes little energy from the mechanical oscillator, Eq. (\ref{Eqfirst}) can be written as \cite{Orszag2016,Zhang2005}:
\begin{equation}
\begin{split}
\frac{d \rho}{dt} \approx \Gamma[\boldsymbol{M}(\tau)-1]\rho  + \mathscr{L}[\rho].
\label{Eqsecond}
\end{split}
\end{equation}
The operator $\mathscr{L}$ in the above equation is attributed to the dissipation of mechanical oscillator and is defined as:
$\mathscr{L}[\rho]\,= \,\frac{\kappa}{2}(\text{n}_{\text{th}}\,+\,1)[2a\rho a^{\dagger}\,-\,a^{\dagger}a\rho\,-\,\rho a^{\dagger}a]\,+\,\frac{\kappa}{2}\text{n}_{\text{th}}[2a^{\dagger} \rho a \,-\,aa^{\dagger}\rho \,-\,\rho aa^{\dagger}]$
with $\kappa$ is the energy decay rate and $\text{n}_{\text{th}}$ is the mechanical oscillator's average phonon number at temperature $\text{T}_{m}$ before cooling and $\text{n}_{\text{th}}\, = \,1/[\exp(\hbar\omega_{\text{m}}/k_{\text{B}}\text{T}_{\text{m}})\,-\,1]$.

With further calculation we can get the mean number of phonon $<\text{n}>_{\text{s}}\,=\,\text{Tr}[a^{\dagger}a\rho_{\text{s}}]$ of the mechanical oscillator in the steady state $\rho_{\text{s}}$ from above master equation:
\begin{align}
\begin{split}
& \frac{d\text{Tr}[a^{\dagger}a\rho_{\text{s}}]}{dt} \\
 = &\Gamma \text{Tr}(a^{\dagger}a[\boldsymbol{M}(\tau)-1]\rho_{\text{s}}) \\
   & +\frac{\kappa}{2}(\text{n}_{\text{th}}+1)\text{Tr}(2a^{\dagger}aa\rho_{\text{s}} a^{\dagger}-a^{\dagger}aa^{\dagger}a\rho_{\text{s}}-a^{\dagger}a\rho_{\text{s}} a^{\dagger}a)\\
   &+\frac{\kappa}{2}\text{n}_{\text{th}}\text{Tr}(2a^{\dagger}aa^{\dagger} \rho_{\text{s}} a -a^{\dagger}aaa^{\dagger}\rho_{\text{s}} -a^{\dagger}a\rho_{\text{s}} aa^{\dagger})\\
= & 0.
\end{split}
\end{align}
Then we get:
\begin{equation}
<\text{n}>_{\text{s}}=\text{n}_{\text{th}}-\frac{\Gamma}{\kappa}\text{Tr}[{a^{\dagger}a(1-\boldsymbol{M}(\tau))\rho_{\text{s}}]}.
\label{mean}
\end{equation}
Here, $\text{Tr}(a^{\dagger}a(1-\boldsymbol{M}(\tau))\rho_{\text{s}})\Gamma/\kappa$ is the decreased phonon number after a $^{87}Rb$ atom is excited from the liquid state to the vapor state. Noticing that
\begin{equation}
\text{Tr}(a^{\dagger}a(1-\boldsymbol{M}(\tau))\rho_{\text{s}})>0,
\end{equation}
can be proved with the definition of the super-operator $\boldsymbol{\hat {M}}(\tau)$. We thus conclude that the temperature of the mechanical oscillator is always cooler than that of its initial state when the steady state is reached. By further calculation we can get that:
\begin{equation}
\text{Tr}(a^{\dagger}a(1-\boldsymbol{M}(\tau))\rho_{\text{s}})=<\text{n}>_{\text{s}}-\text{Tr}[a^{\dagger}a\boldsymbol{M}(\tau)\rho_{\text{s}}].
\label{steady}
\end{equation}

Substituting Eq. (\ref{trace}) into Eq. (\ref{steady}), and the second term of the right side of Eq. (\ref{steady}) can be expanded as:
$\text{Tr}[a^{\dagger}a\boldsymbol{M}(\tau)\rho_{\text{s}}]\,=\,\text{Tr}[a^{\dagger}a(\bra{0}e^{-iH_{I}\tau/\hbar}\rho_{\text{s}}\otimes\ket{0}\bra{0}e^{iH_{I}\tau/\hbar}\ket{0}
+\bra{1}e^{-iH_{I}\tau/\hbar}\rho_{\text{s}}\otimes\ket{0}\bra{0}e^{iH_{I}\tau/\hbar}\ket{1})].$  Here, we proceed by noting that for small time $\tau$, with $g_{0}\tau/2<1$ which can be proved in our further calculation with some practical parameters, and we get:
\begin{equation}
\begin{split}
\text{Tr}\{a^{\dagger}a[1-\boldsymbol{M}(\tau)]\rho_{\text{s}}\}&=\text{Tr}[a^{\dagger}a\rho_{\text{s}}]-\text{Tr}[a^{\dagger}a\boldsymbol{M}(\tau)\rho_{\text{s}}]\\
&\approx\frac{1}{8} g_{0}^{2}\tau^{2}<\text{n}>_{\text{s}}.
\label{Eqdecreasing}
\end{split}
\end{equation}

Substituting Eq. (\ref{Eqdecreasing}) into Eq. (\ref{mean}), the mean steady-state phonon number after cooling is given by:
\begin{equation}
<\text{n}>_{\text{s}}=\frac{\text{n}_{\text{th}}}{1+\frac{1}{8}g^{2}_{0}\tau^{2}\frac{\Gamma}{\kappa}}.
\label{Eqsingle}
\end{equation}
When there are N BEC atoms in the magnetic trap which are identically coupled to the mechanical oscillator during the cooling process, the coupling constant is collectively enhanced \cite{Treutlein2007,Aspelmeyer2014}: $g_{0}\,\rightarrow\,g_{0}\sqrt{N}\,=\, g_{N}$. Therefore, for N cold atoms coupling the mechanical oscillator, the mean  steady-state phonon number is given by:
\begin{equation}
<\text{n}>_{\text{s}}=\frac{\text{n}_{\text{th}}}{1+\frac{1}{8}g_{N}^{2}\tau^{2}\frac{\Gamma}{\kappa}}.
\label{Eqcollective}
\end{equation}
Combining $g_{N}$, $g_{0}$,  $\kappa\,=\,\text{Q}_{\text{m}}/\omega_{\text{m}}$, and Eq. (\ref{time}) into Eq. (\ref{Eqcollective}) we have:
\begin{align}
\begin{split}
<\text{n}>_{\text{s}}
&=\frac{\text{n}_{\text{th}}}{1+(\frac{\pi\mu_{B}}{8\hbar})^{2}\, N\,\zeta(\delta)\,(G_{\text{m}}a_{\text{qm}})^{2}\,\frac{\text{Q}_{\text{m}}}{\omega_{\text{m}}}}.
\label{expression}
\end{split}
\end{align}
Equation (\ref{expression}) gives the mean steady-state phonon number after cooling of the mechanical oscillator, with $\text{Q}_{\text{m}}$ being the quality factor of the mechanical oscillator. In the following we will discuss the mean steady-state phonon number after cooling with some practical parameters.

\section{discussion and conclusion}
\label{discussion and conclusion}

\begin{figure}[bp]
\centering
\includegraphics[width=8.5cm]{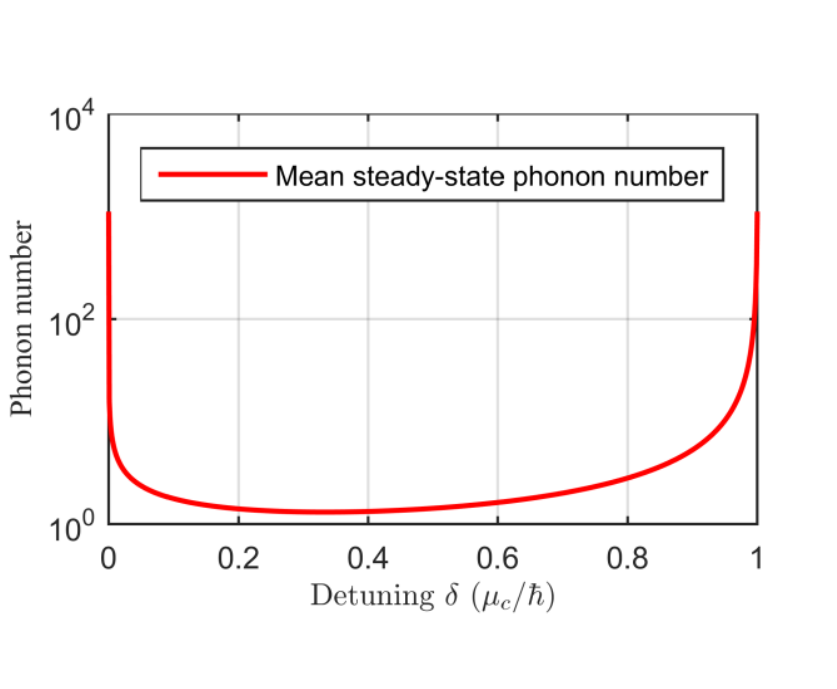}
\caption{(Color online). Steady phonon number of mechanical oscillator after cooling depends on the detuning with the initial temperature of mechanical oscillator is 50 mK.}
\label{fig:3}
\end{figure}

We now check the cooling limit with some practical parameters. Let us consider a mechanical oscillator realized by a silicon cantilever: $\omega_{\text{m}}/2\pi\,=\,1.0$ MHz, the quality factor $\text{Q}_{\text{m}}\,=\,1\,\times\,10^{5}$ and effective mass $m_{\text{eff}}\,=\,10^{-16}$ kg, so that the zero point fluctuation $a_{\text{qm}}\, = \,2.9\,\times\,10^{-13}$ m. If the initial temperature of this mechanical oscillator before cooling is 50 mK, the amplitude of the thermal fluctuation of mechanical oscillator is about $1.86\,\times\,10^{-10}$ m. Therefore, the assumption of $ g_{0}\tau/2\, < \,1$ is justified. Let the single-atom-single-phonon coupling constant $g_{0}\,=\,8$ Hz which can be realized by adjusting the magnetic tip \cite{ Treutlein2007,Tretiakov2016}. The energy decay rate for this mechanical oscillator is: $\kappa\,=\,\omega_{\text{m}}/\text{Q}_{\text{m}}\,=\,2\pi\,\times\,10$ Hz . The chemical potential $\mu_{c}\,\propto\,(N\omega_{\text{x}}\omega_{\text{y}}\omega_{\text{z}})^{2/5}$ \cite{Darazs2014}. For the frequencies of the trap: $\omega_{\text{x}}/2\pi\,=\,\omega_{\text{y}}/2\pi\,=\,250$ Hz and $\omega_{\text{z}}/2\pi\,=\,19$ Hz with the number of atoms $\text{N}=5\times10^{6}$ \cite{Steck1998}, we have $\mu_{c}/\hbar\,\approx\,2\pi\,\times\,2.88$ kHz, so that $\zeta(\delta)\,=\,15\pi\hbar\,\times\,[\sqrt{\hbar\delta/\mu_{c}}\,-\,(\sqrt{\hbar\delta/\mu_{c}})^{3}]/8\mu_{c}\,\approx\,3.3\,\times\,10^{-4}\,\times\,[\sqrt{\hbar\delta/\mu_{c}}\,-\,(\sqrt{\hbar\delta/\mu_{c}})^{3}]$. With these parameters, we can calculate the mean steady-state phonon number $<n>_{\text{s}}$ of mechanical oscillator after cooling. The phonon number is depicted in Fig. \ref{fig:3} which is depending on detuning $\delta$.

The mean steady-state phonon number after cooling is shown in Fig. \ref{fig:3} with the initial temperature before cooling is 50 mK. The phonon number depends on detuning $\delta$ is expressed in units of $\hbar/\mu_{c}$. Firstly, the mean steady-state phonon number decreases apparently as the increasing of detuning $\delta$ before $\hbar\delta/\mu_{c}\,=\,0.2$. When $\hbar\delta/\mu_{c}$ is in the range of 0.2 - 0.6, the steady-state phonon number change slowly with the change of the detuning. It means that the mechanical oscillator can be cooled robustly which allows a wide range to adjust the detuning. The mean phonon number in steady state reaches the minimum when $\hbar\delta/\mu_{c}\,=\,1/3$, we will call it as best point in the following. The steady-state phonon number of mechanical oscillator after cooling at best point is $<\text{n}>_{\text{s}}\,\approx\,1.3$.

\begin{figure}[bp]
\centering
\includegraphics[width=8.5cm]{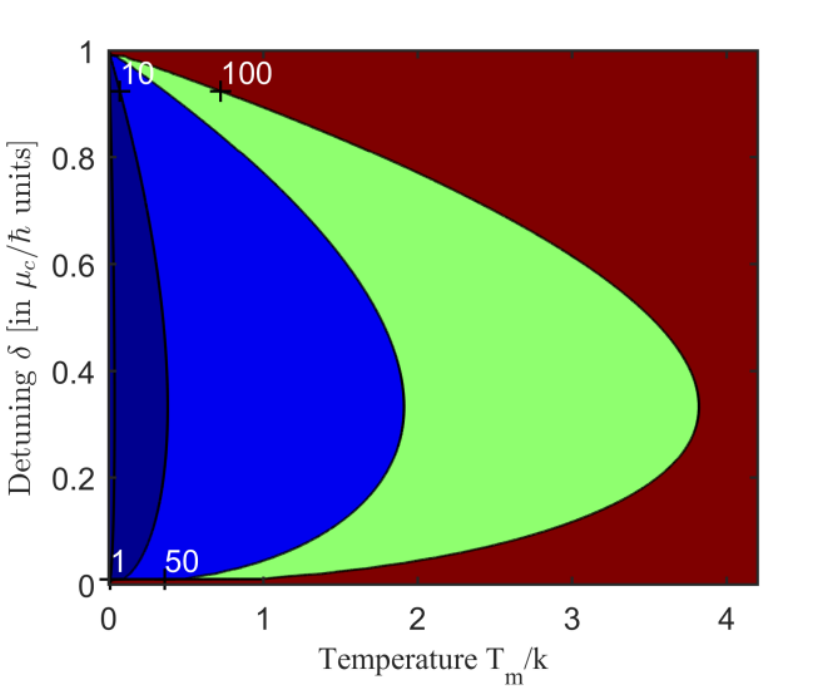}
\caption{(Color online). Mean steady-state phonon number of mechanical oscillator after cooling, the initial temperature we considering is in the range of 0 - 4.2 K.}
\label{fig:4}
\end{figure}

We then explored the a wide range of initial temperature for the mechanical oscillator. The temperature is in the range of 0 - 4.2 K. The mechanical oscillator with $\omega_\text{m}/2\pi\,=\,1.0$ MHz, $\text{Q}_{\text{m}}\,=\,1\,\times\,10^5$ and effective mass $m_{\text{eff}}\,=\,10^{-16}$ kg. The range of initial temperature of this mechanical oscillator in which the oscillator can be cooled to the ground state is depicted in Fig. \ref{fig:4}, with mean steady-state phonon number less than 1. We find that the maximum initial temperature that can be cooled to ground state is about 40 mK if the detuning is set at the best point. This temperature can be realized if the oscillator is placed in a dilution refrigerator.

\begin{figure}[tp]
\centering
\includegraphics[width=8.5cm]{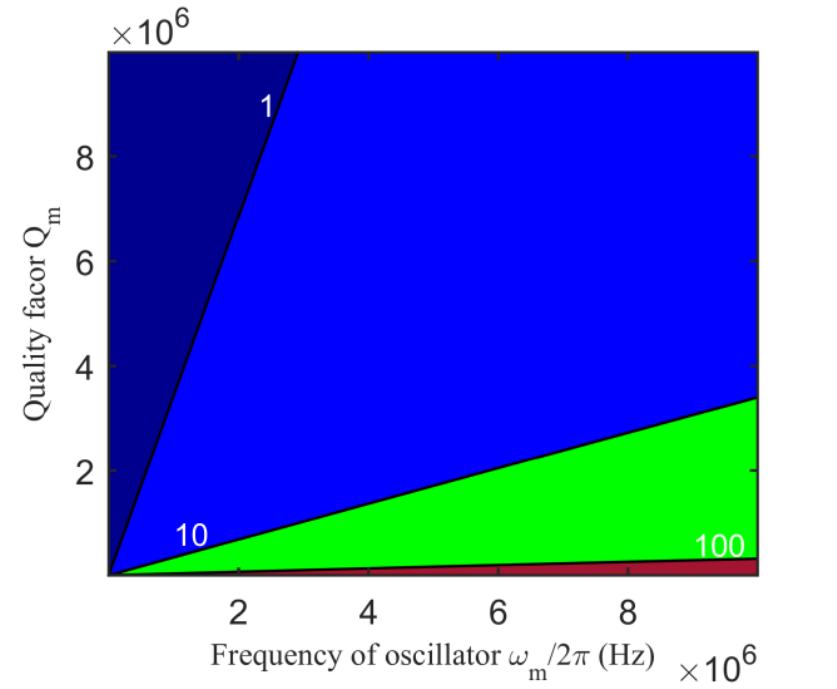}
\caption{(Color online). Mean steady-state phonon number of mechanical oscillator after cooling. The quality factor $\text{Q}_{\text{m}}$ is in the range of $10^{3}\,-\,10^{7}$ and frequency $\omega_{\text{m}}/2\pi$ is in the range of $10^{3}\,\text{Hz}\,-\,10^{7}\,\text{Hz}$. The initial temperature for these mechanical oscillators before cooling is $\text{T}_{\text{m}}\,=\,4.2$ K.}
\label{fig.5}
\end{figure}

We also exploring the range of quality factor $\text{Q}_{\text{m}}$ and frequency of mechanical oscillator $\omega_{\text{m}}$ when it is at the initial temperature 4.2 K in Fig. \ref{fig.5}. We can find that when the quality factor $\text{Q}_{\text{m}}$ is about $1\,\times\,10^{5}$, the ground state can be reached even the initial temperature $\text{T}_{\text{m}}$ is at the 4.2 K, when the frequency of mechanical oscillator $\omega_{\text{m}}/2\pi\,=\,1$ kHz. When the mechanical oscillator is at the initial temperature of 4.2 K, we find that the range of $\text{Q}_{\text{m}}$ and $\omega_{\text{m}}$ is shown at the top left corner in Fig. \ref{fig.5}. What is more, there is a wide range of mechanical oscillators which can be cooled to a temperature with only few phonons, even the initial temperature is 4.2 K. It means that, the inequality $Q_{\text{m}}\,\times\,f_{\text{m}}\,>\,k_{\text{B}}T_{\text{m}}/h$, the measure of an experiment for which could reached into quantum regime for the manipulation of coherent phonon \cite{Norte2016}, could be well satisfied, even for the bad quality oscillators. Here, $f_{\text{m}}\,=\,\omega_{\text{m}}/2\pi$, is the frequency of the mechanical oscillator.

We can come to the conclusion that, our cooling method with the BEC state cooling mechanical oscillator is an effective proposal which can cool the mechanical oscillator to the ground state at high initial temperature range. If the Cobalt bar is replaced by a Dysprosium bar, which has a higher value of magnetic momentum $\mu_{\text{m}}$ than Cobalt bar in unit mass, the value of single-atom-single-phonon coupling will be bigger. In this way, the mechanical oscillator will be colder than the case of Cobalt bar, and the ground state will be easier to be reached.

In summary, we studied a cooling proposal for mechanical oscillator. We find that by applying this method, the mechanical oscillator can be cooled to the ground state. We harbor the idea that that this cooling method can be demonstrated in the near future by combining rapid developing techniques for micro trap on atom chips and nano-mechanical oscillators.

\section{Acknowledgments}
D.H.X. thanks Yang Yu for his help on drawing of graphs. This work was supported by the Technological Development Grant of Hefei Science Center of Chinese Academy of Sciences, Grants 2014TDG-HSC001; the National Key Research and Development Program of China, Grants 2017YFA0303201; F. X. also thanks the support of  Recruitment Program for Young Professionals.

\newpage


\begin{thebibliography}{99}

\bibitem{Moser2013}
J. Moser, J. G$\ddot{u}$ttinger, A. Eichler, M. J. Esplandiu, D. E. Liu, M. I. Dykman and A. Bachtold, Nature Nanotechnology \textbf{8},493-496 (2013)

\bibitem{Chaste2012}
J. Chaste, A. Eichler, J. Moser, G. Ceballos, R. Rurali and A. Bachtold, Nature Nanotechnology \textbf{7}, 301-304 (2012).


\bibitem{Xue2011}
F. Xue, D. P. Weber, P. Peddibhotla, and M. Poggio, Phys. Rev. B \textbf{84}, 205328 (2011).

\bibitem{Peddibhotla2013}
P. Peddibhotla, F. Xue, H. I. T. Hauge, S. Assali, E. P. A. M. Bakkers, and M. Poggio, Nature Physics \textbf{9}, 631-635 (2013).

\bibitem{Cleland2004}
A. N. Cleland and M. R. Geller, Phys. Rev. Lett. \textbf{93}, 070501 (2004).

\bibitem{Wang2011}
H. Wang, Matteo Mariantoni, Radoslaw C. Bialczak, M. Lenander, Erik Lucero, M. Neeley, A. D. O'Connell, D. Sank, M. Weides, J. Wenner, T. Yamamoto, Y. Yin, J. Zhao, John M. Martinis, and A. N. Cleland,
Phys. Rev. Lett. \textbf{106}, 060401 (2011).

\bibitem{Xue2007a}
F. Xue, Y. D. Wang, C P Sun, H Okamoto, H Yamaguchi and K Semba, New J. Phys. \textbf{9} 35 (2007).

\bibitem{Rabl2010}
P. Rabl, S. J. Kolkowitz, F. H. L. Koppens, J. G. E. Harris, P. Zoller and M. D. Lukin, Nature Phys. \textbf{6}, 602 (2010).

\bibitem{Connell2010}
A. D. O'Connell, M. Hofheinz1, M. Ansmann, Radoslaw C. Bialczak, M. Lenander, Erik Lucero, M. Neeley1, D. Sank, H. Wang, M. Weides, J. Wenner, John M. Martinis, and A. N. Cleland, Nature(London), \textbf{464}: 697-703 (2010).


\bibitem{Aspelmeyer2014}
M. Aspelmeyer, T. J. Kippenberg, and F. Marquardt, Rev. Mod. Phys. \textbf{86}, 1391 (2014).

\bibitem{Naik2006}
A. Naik, O. Buu, M. D. LaHaye, A. D. Armour, A. A. Clerk, M. P. Blencowe, and K. C. Schwab, Nature (London), \textbf{443}(7108): 193-196 (2006).

\bibitem{Teufel2008}
J D Teufel, C A Regal, and K W Lehnert, New J. Phys. \textbf{10} 095002 (2008).

\bibitem{Tian2009}
L. Tian Phys. Rev. B \textbf{79}, 193407 (2009) .

\bibitem{Xue2007b}
F. Xue, Y. D. Wang, Yu-xi Liu, and Franco Nori, Phys. Rev. B \textbf{76}, 205302 (2007).

\bibitem{Marquardt2007}
F. Marquardt, Joe P. Chen, A. A. Clerk, and S. M. Girvin, Phys. Rev. Lett. \textbf{99}, 093902 (2007).

\bibitem{Treutlein2007}
P. Treutlein, D. Hunger, S. Camerer, T. W. H$\ddot{a}$nsch, and J. Reichel, Phys. Rev. Lett. \textbf{99}, 140403 (2007).

\bibitem{Deng2014}
Y. Deng, J. Cheng, H. Jing, and S. Yi, Phys. Rev. Lett. \textbf{112}, 143007 (2014)

\bibitem{Tretiakov2016}
A. Tretiakov, and L. J. LeBlanc, Phys. Rev. A \textbf{94}, 043802 (2016).

\bibitem{Wang2006}
Y. J. Wang, M. Eardley, S. Knappe, J. Moreland, L. Hollberg, and John Kitching, Phys. Rev. Lett. \textbf{97}, 227602 (2006).

\bibitem{Hunger2010}
D. Hunger, S. Camerer, T. W. H$\ddot{a}$nsch, D. K$\ddot{o}$nig, J. P. Kotthaus, J. Reichel, and P. Treutlein, Phys. Rev. Lett. \textbf{104}, 143002 (2010).

\bibitem{Hunger2011}
D. Hunger, S.Camerer, M.Korppi, A. J$\ddot{o}$ckel, T.W.H$\ddot{a}$nsch, and P.Treutlein, Comptes Rendus Physique, \textbf{12}(9), 871-887 (2011).

\bibitem{Montoya2015}
C. Montoya, J. Valencia, A. A. Geraci, M. Eardley, J. Moreland, L. Hollberg, and J. Kitching, Phys. Rev. A \textbf{91}, 063835 (2015).

\bibitem{Leggett2001}
A. J. Leggett, Rev. Mod. Phys. \textbf{73}, 307 (2001).

\bibitem{Mewes1997}
M.-O. Mewes, M. R. Andrews, D. M. Kurn, D. S. Durfee, C. G. Townsend, and W. Ketterle, Phys. Rev. Lett. \textbf{78}, 582 (1997)

\bibitem{Steck1998}
H. Steck, M. Naraschewski, and H. Wallis, Phys. Rev. Lett. \textbf{80}, 1 (1998).

\bibitem{Patton2013}
Kelly R. Patton, and Uwe R. Fischer, Phys. Rev. Lett. \textbf{111}, 240504 (2013).

\bibitem{Steinke2011}
S. K. Steinke, S. Singh, M. E. Tasgin, P. Meystre, K. C. Schwab, and M. Vengalattore, Phys. Rev. A \textbf{84}, 023841 (2011).

\bibitem{Darazs2014}
Z. Dar$\acute{a}$zs, Z. Kurucz, O. K$\acute{a}$lm$\acute{a}$n, T. Kiss,  J. Fort$\acute{a}$gh, and P. Domokos, Phys. Rev. Lett. \textbf{112}, 133603 (2014).

\bibitem{Woodgate1980}
G. K. Woodgate. Elementary atomic structure (1980).

\bibitem{Treutlein2006}
P. Treutlein, T. W. H$\ddot{a}$nsch, J. Reichel, A. Negretti, M. A. Cirone, and T. Calarco, Phys. Rev. A \textbf{74}, 022312 (2006).

\bibitem{Geraci2009}
Andrew A. Geraci and John Kitching, Phys. Rev. A \textbf{80}, 032317 (2009).

\bibitem{Leggett2006}
A. J. Leggett, \emph{Quantum liquids: Bose condensation and Cooper pairing in condensed-matter systems} (Oxford University Press, 2006) Chap. 4, The Bose alkali gases.

\bibitem{Shore1993}
Bruce W. Shore and Peter L. Knight, Journal of Modern Optics, \textbf{40} (7): 1195-1238 (1993).

\bibitem{Xichen2015}
Xi Chen, Yong-Chun Liu, Pai Peng, Yanyan Zhi, and Yun-Feng Xiao, Phys. Rev. A \textbf{92}, 033841 (2015).


\bibitem{Orszag2016}
M. Orszag, \emph{Quantum Optics: Including Noise Reduction,Trapped Ions, Quantum Trajectories, and
Decoherence}(Springer, Berlin, 2016), Chap. 11, Quantum Laser Theory: Master Equation Approach.

\bibitem{Scully1997}
M. O. Scully, and M. S. Zubairy, \emph{Quantum Optics}(Cambridge University Press, 1997), Chap.8, Quantum Theory of Damping-Density Operator and Wave Function Approach.

\bibitem{Zhang2005}
P. Zhang, Y. D. Wang, and C. P. Sun Phys. Rev. Lett. \textbf{95}, 097204 (2005)

\bibitem{Norte2016}
R.A. Norte, J.P. Moura, and S. Gr$\ddot{o}$blacher Phys. Rev. Lett. \textbf{116}, 147202 (2016)

\end{thebibliography}
\end{document}